 \documentclass[12pt,preprint]{aastex}

 \usepackage{natbib}

 \slugcomment{will appear in 2007 ApJ, 656 (in press)}

 \shorttitle{Determination of the Coronal Magnetic Field by Hot Loop Oscillations}
 \shortauthors{Wang et al.}

\begin{document}

 \title{Determination of the Coronal Magnetic Field by Hot Loop Oscillations Observed by
  SUMER and SXT}

 \author{Tongjiang Wang\altaffilmark{1}, Davina E. Innes\altaffilmark{2},
   and Jiong Qiu\altaffilmark{1}}

 \altaffiltext{1}{Department of Physics, Montana State University, Bozeman, MT 59717-3840}
 \altaffiltext{2}{Max-Planck-Institut f\"{u}r Sonnensystemforschung,
     D-37191 Katlenburg-Lindau, Germany}

 \begin{abstract}
We apply a new method to determine the magnetic field in coronal loops using
observations of coronal loop oscillations. We analyze seven Doppler shift
oscillation events detected by SUMER in the hot flare line Fe~{\small XIX}
to obtain oscillation periods of these events. The geometry, temperature, and electron 
density of the oscillating loops are measured from coordinated multi-channel soft X-ray 
imaging observations from SXT. All the oscillations are consistent with standing slow 
waves in their fundamental mode. The parameters are used to calculate the magnetic field 
of coronal loops based on MHD wave theory. For the seven events, the plasma
$\beta$ is in the range  0.15$-$0.91 with a mean of 0.33$\pm$0.26, and the estimated
magnetic field varies between 21$-$61 G with a mean of 34$\pm$14 G. With background
emission subtracted, the estimated magnetic field is reduced by 9\%$-$35\%.  The
maximum backgroud subtraction gives a mean of 22$\pm$13~G in the range 12$-$51~G. We discuss 
measurement uncertainties and the prospect of determining coronal loop magnetic fields 
from future observations of coronal loops and Doppler shift oscillations.
\end{abstract}


\section{Introduction}
Magnetic field governs the structure and dynamics of the Sun's corona.
Although the structure and topology of the coronal magnetic field can be indirectly traced in
imaging observations of coronal loops in X-rays and EUVs, a direct measurement of the coronal magnetic field
remains a very difficult and challenging problem. In the past, efforts have been made with microwave,
optical and EUV observations to measure coronal magnetic fields.
The microwave gyroresonance magnetometry technique is used to measure strong (more than a few hundred gauss)
active region field strengths in the low corona \citep[e.g.][]{gar94, sch94, lee99},
and  Faraday rotation observations of polarized radiation by radio sources can be used to measure the
mean line-of-sight coronal field \citep[e.g.][]{sak94}. At optical and EUV wavelengths, the Zeeman
and Hanle effects are used to determine the coronal magnetic field, but their applications are limited
because the signals are weak. Recently, \citet{lin00, lin04}
accurately measured the line-of-sight component of the coronal magnetic field using Zeeman splitting
observations in the near-infrared line Fe~{\small XIII} 10747 \AA~and obtained a field strength of 4$-$33~G
at a height ranging between 0.10 and 0.15 of a solar radius off the limb above active regions. \citet{rao02} 
made the first attempt to measure the coronal field from the linear polarization of the O~{\small VI}
1032 \AA~line due to the Hanle effect and deduced a coronal field of about 3 G in a polar hole.

A new methodology, known as {\em MHD coronal seismology} \citep[e.g.][]{rob84,rob04},
has been developed to diagnose plasma properties of coronal loops exhibiting oscillation patterns.
\citet{nak01} first applied this method to TRACE observations of global fast kink-mode loop oscillations
in two events, yielding a loop field strength in the range of 4$-$30~G. Taking into consideration
the plasma density in one event, they refined their determination to 13$\pm$9 G. With the same method,
\citet{asc02} obtained the magnetic field strength ranging between 3 and 90~G for 26 oscillating loops
observed by TRACE. 

In this paper we analyze slow mode MHD waves in coronal loops to determine their magnetic field. Standing 
slow mode waves, triggered by small flares, were recently discovered in spectroscopic observations taken 
by SUMER on-board SOHO, by their Doppler shift signatures in the flare lines Fe {\small XIX} and Fe {\small XXI}
\citep[e.g][]{wan02, wan03a, wan03b}. Similar strongly damped Doppler shift oscillations were also detected
by BCS on {\em Yohkoh} in many flares, which were also interpreted as signatures of standing slow mode
waves \citep{mar05, mar06}. The plasma $\beta$ in hot ($>$ 6 MK) loops can be much larger than the expected 
mean value ($\sim$0.01) in the corona, which implies a relatively strong coupling between the magnetic 
and thermal perturbations. Therefore, the loop field strength may also be deduced from diagnostics 
of standing slow mode oscillations in hot coronal loops \citep{rob04}. In this study we will test 
this idea. In the following text, the observations are described in Sect.~2. The method for deducing 
the magnetic field strength is given in Sect.~3. We present the results and uncertainty analysis 
in Sect.~4, and the conclusions and discussions in Sect.~5.

\section{Observations}
Spectral observations of active region coronae were obtained by SUMER
in sit-and-stare mode. Two neighboring active regions on the east limb were observed with
a cadence of 90~s using the $300{''}\times4{''}$ slit during 2000 September 16--20.
Five spectral lines, Fe~{\small XIX} $\lambda$\,1118 (6.3 MK), Ca~{\small XV}
$\lambda$\,1098 and $\lambda$\,555$\times$2 (3.5 MK), Ca~{\small XIII}
$\lambda$\,1134 (2.2 MK), and Si~{\small III} $\lambda$\,1113 (0.06 MK) were transmitted, with a
2.2\AA~wide window for each line. After processing the raw data following standard procedures
(decompression and corrections of flat-field, detector distortions, deadtime, and gain effects),
a single Gaussian was fit to each line profile to obtain a Doppler shift time series at each
spatial pixel (e.g. Fig.~\ref{fig1}{\em c}). A large number of flarelike brightenings were revealed in
the hot flare line, Fe {\small XIX}. They have an occurrence rate 3$-$14 per hour and their lifetimes
range from 5$-$150 min, with an average of about 25 min \citep{wan06}. From about 180 events identified by
the visual inspection, more than 20\% of them are found to be associated with Doppler shift
oscillations.

For six Doppler shift events in this dataset, coordinated imaging observations were obtained by SXT on {\em Yohkoh}.
In addition, coordinated SXT observations were also made for an oscillation event reported in an earlier
study by \citet{wan02}, thus this event is also included in this study.  Analyses of joint SXT and SUMER
observations of coronal loop oscillations allow us to identify the oscillating loops, determine their
geometric parameters, and derive their temperature and density.
In order to compare SXT and SUMER observations, the scale of the SXT full (2${''}$.5~pixel$^{-1}$)
and half-resolution images is increased by a factor of about 1\% to account for the different orbits of
{\em Yohkoh} and SOHO. Then the SXT images are coaligned with the SUMER slit according to their pointing information.
The Fe~{\small XIX} brightenings along the slit coincide accurately with the slit intersection on the SXT loops,
indicative of a good coalignment (see Figs.~\ref{fig1}-\ref{fig5}{\em a} and~\ref{fig1}-\ref{fig5}{\em b}).
The average temperature and emission measure of the loops are derived from the SXT images using the filter ratio
method implemented in the standard SolarSoftware under the assumption of isothermal plasma \citep{tsu91, har92}.
Only the images without saturated pixels in the region of oscillating loops are used.

\section{Method for determining the magnetic field strength based on global standing slow mode
oscillations}
  \label{secmet}

Applying the MHD wave theory for a straight magnetic cylindrical model to the coronal loop
 \citep{edw83, rob84}, the oscillation period of a standing slow mode wave
in its fundamental mode is given by,
\begin{equation}
 P=\frac{2L}{c_t}, \hspace{3mm} c_t=\left(\frac{1}{c_s^2}+\frac{1}{v_A^2}\right)^{-1/2}, \label{eqpslw}
\end{equation}
\noindent
where $L$ is the loop length, $c_s=(\gamma{k_B}T/\mu{m}_p)^{1/2}=1.52\times10^4T^{1/2}$
cm~s$^{-1}$ is the sound speed, and $v_A=B/(\mu_0\rho)^{1/2}=2.18\times10^{11}Bn_e^{-1/2}$ cm~s$^{-1}$
is the Alfv\'{e}n speed. $T$ is the plasma temperature (K), $B$ the magnetic field strength (G) in the loop, and
$n_e$ the electron density (cm$^{-3})$. Finally, $\gamma$ is taken as 5/3, $k_B$ the Boltzmann constant,
$m_p$ the mass of a proton and $\mu$ the mean molecular weight, taken here to be 0.6 assuming
a standard coronal He abundance.

A recent work of \citet{rob06} shows that the stratification will modify the period of slow modes by,
\begin{equation}
 P=\frac{2L}{c_t}\cdot\frac{1}{{(1+L^2/L_c^2)}^{1/2}}, \hspace{3mm} L_c\approx 2\pi\Lambda, \label{eqmp}
\end{equation}
\noindent
where $\Lambda=k_B T/\mu m_p g\approx50$ Mm ($T$/1.0 MK) is the pressure scale height. For hot loops of
T=6 MK, it is shown that $L_c=1900$ Mm, far greater than the observed typical loop length. Hence
gravitational effects on the period, $P$, of slow modes in hot SUMER loops are negligible.

From equation~(\ref{eqpslw}) we have
\begin{equation}
  P=\frac{2L}{c_t}=\frac{2L}{c_s}\left(1+\frac{\gamma}{2}\beta \right)^{1/2}, \hspace{3mm}
  \beta=\left(\frac{2}{\gamma}\right)\frac{c_s^2}{v_A^2}.  \label{eqbt}
\end{equation}
This equation shows that when the plasma $\beta$ is small, the oscillation period
is essentially independent of the magnetic field, which is normally the case in the corona.
Then  the derivation of the magnetic field will be very sensitive to
uncertainties in period, loop length and sound speed measurements. However, in
the case of hot loops, $\beta$ is not so small. If we assume a coronal loop with an electron density of
$10^9$ cm$^{-3}$ and a magnetic field of 10 G, then $\beta$=0.06 for cool (1 MK) loops while
$\beta$= 0.35 for hot (6 MK) loops. Thus it is reasonable to apply this method to hot loops.
From equation~(\ref{eqpslw}) or~(\ref{eqbt}), the magnetic field can be derived by
\begin{equation}
B=\left(\frac{n_9}{C_1}\right)^{1/2}\left(\frac{P^2}{4L^2}-\frac{1}{C_2T_6}\right)^{-1/2},  \label{eqmb}
\end{equation}
\noindent
where $C_1=4.8\times10^3$ and $C_2=2.3\times10^4$ are constants. $B$, $n_9$, $T_6$, $P$, and $L$
are in units of G, $10^9$~cm$^{-3}$, MK, s and km, respectively. When $n_9$, $T_6$, $P$, and $L$
are measured from the coordinated SUMER and SXT observations, $B$ can be determined using 
equation~(\ref{eqmb}).

\section{Results}
\subsection{SUMER Observations of Hot Loop Oscillations}
\label{subsum}
We first analyze the Doppler shift oscillations seen by SUMER with the method used in
our previous studies \citep[e.g.][]{wan02,wan03a,wan03b}. Figures~\ref{fig1}-\ref{fig6}{\em b}
and~\ref{fig1}-\ref{fig6}{\em c} show the
time series of intensity and Doppler shift in the Fe {\small XIX} line. We find that
in all cases the regions that exhibited prominent Doppler shift oscillations are located at the loop
top. In contrast, when the SUMER slit intersected with the loop legs, no noticeable
evidence for Doppler shift oscillations
is found. This is shown in Fig.~\ref{fig7}. The southern loop system is the same as that shown in
Fig.~\ref{fig1}, but closer to the limb. Two intensity peaks occurred simultaneously at the legs
where the slit intersected with the loop. The time profile of Doppler shifts reveals flows in the loop,
and only very weak oscillations. Figure~\ref{fig2} shows another example. In this case, the slit intersected one
leg of a large loop in the north of the analyzed oscillating loop (Fig.~\ref{fig2}{\em a}). Similarly,
the evolution of Doppler shifts shows flows (up to 40 km s$^{-1}$) in the loop, but without oscillations.
The dominance of the oscillation at the loop top over the leg regions is consistent with the
features of a standing slow wave in its fundamental mode, which has an anti-node at the loop apex
and nodes at the footpoints of the loop in velocity profiles.

Recent MHD simulations have shown that both the fundamental mode \citep[e.g.][]{sel05,tar05} and
the second harmonic mode \citep[e.g.][]{nak04,tsi04} of standing slow waves can be generated in 
a coronal loop. But so far no evidence for the second harmonic has been found in SUMER observations
\citep{wan03b}. The period of the second harmonic is  half the period of the fundamental mode and
its velocity oscillation has a node at the apex and two anti-nodes at the legs of the loop. Therefore,
strong (weak) anti-phase Doppler shift oscillations would be expected along the two legs (the top).
We note that in three cases (Loop-1 shown in Fig.~\ref{fig1}, Loop-3 in Fig.~\ref{fig2}, and Loop-5
in Fig.~\ref{fig4}) the loop top brightenings seem to be associated with Doppler shift oscillations
in anti-phase, suggesting the possibility of the second harmonics. However, we find that the measured
oscillation periods are consistent with those estimated for the fundamental mode, which are approximately
twice the sound wave transit time along the loop given the sound speed listed in
Table~{\ref{tabpar} (see Sect.~\ref{sectsxt}). Therefore, this signature
is probably related to contamination by oscillations from nearby (or super-imposed) loops. For example,
an arcade system consists of several loops with similar lengths, but their oscillations are excited
in anti-phase. This explanation is well supported by the case of Loop-1 (the northern loop system
shown in Fig.~\ref{fig1}{\em a}). We find that the southern part of the Doppler shift oscillations is
more regular and coherent along the slit, while the northern part is complicated (Fig.~\ref{fig1}{\em c}).
The evolution of intensity along the slit reveals two close brightenings (Fig.~\ref{fig1}{\em b}),
which correspond to different parts of the oscillations, respectively. In addition, Figure~\ref{fig7}
also supports that Loop-1 is a loop (or arcade) system. In the case of Loop-3, the two parts of 
the oscillations show different decay times. The southern part is visible for about five
periods, while the northern part is only visible for about two periods (Fig.~\ref{fig2}{\em c}). In the case of
Loop-5, the two neighboring oscillations are not exactly in anti-phase. The oscillation period of the southern
one is smaller than the northern one (Fig.~\ref{fig4}{\em c}). Therefore, events studied in this
paper are all considered to exhibit fundamental modes.

We measure physical parameters of the oscillations averaged along the slit in a region where the
Doppler shift oscillations are strongest (the cuts marked with two parallel lines in panel ({\em c}) of
each figure). For Loop-1, Loop-3 and Loop-5, we choose the regions of more regular and
coherent oscillations. The function
\begin{equation}
V(t)=V_{0}+V_{\rm m}{\rm sin}(\omega{t}+\phi)e^{-\lambda{t}},
\label{fitequ}
\end{equation}
\noindent
was then fit to the oscillation (Figs.~\ref{fig1}-\ref{fig6}{\em d}), where $V_{0}$ is the background
Doppler shift, $V_{\rm m}$ is the Doppler shift amplitude and  $\omega$, $\phi$, and $\lambda$ are
the frequency, phase, and decay rate of the oscillations, respectively. We derive the maximum displacement
amplitude by $A=V_{\rm m}/(\omega^2+\lambda^2)^{1/2}$. The obtained parameters of the time series
are listed in Table~\ref{tabosc}. We find that the oscillation periods in the range of 8$-$18 min and
the decay times of 7$-$20 min are all consistent with the results by \citet{wan03b} in a previous
study of 54 oscillation events. The oscillation events in Loop-3 and Loop-5 are unusual because they are
visible for more than 5 periods (Fig.~\ref{fig2}{\em d} and~\ref{fig4}{\em d}). Loop-3 exhibits the
slowest decay we have ever found \citep{wan03a}, with a ratio of the decay time to the period being about 2.4.
It was associated with a C3.1 flare.

\subsection{SXT Observations of Oscillating Loops}
\label{sectsxt}
We measure geometrical parameters of the oscillating loops seen in SXT images, especially the
loop length, which is a critical parameter to identify the oscillation mode and evaluate
the magnetic field in coronal loops as will be shown in Sect.~\ref{subcmf}.
We use a method similar to that employed by \citet{asc02}. By assuming a circular loop shape, this
method optimizes two free parameters ($h_0$ and $\theta$) to obtain a good match with the observed
loop, where $h_0$ is the height of the circular loop center above the solar surface and $\theta$ is
the inclination angle of the loop plane to the vertical. This is shown in Fig.~\ref{fig1}{\em a}.
We notice that two loops, named Loop-2a and Loop-2b, can be discerned in the southern loop system.
For Loop-2b, an elliptical shape must be assumed in order to get the best fit, in which an additional
free parameter of ellipticity, $e$, is included. For Loops 4$-$7 which were just located 
above the limb, because of an uncertainty in determining the positions of their footpoints, we
apply another method described by \citet{wan03b} in Appendix A. Based on a
circular model, the geometric parameters of the loop can be derived from measurements of the
limb footpoint separation and the apex position of the loop by optimizing the free parameter $h_0$.
Note that with this method the analyzed loop is assumed to have the midpoint between its footpoints
exactly above the limb. The obtained loop length, the inclination angle and the azimuth angle of
the footpoint baseline are listed in Table~\ref{tabgeo}.

We compare the evolution of intensity measured at the loop apex by SUMER and SXT.
Figures~\ref{fig1}-\ref{fig6}{\em e} show that the time profiles of soft X-ray flux in the SXT AlMgMn
filter and the SUMER Fe~{\small XIX} intensity coincide well, indicating that the
Fe~{\small XIX} observations are a good reflection of the coronal soft X-ray emission. In our
previous studies \citep[e.g.][]{wan02,wan03a,wan03b}, we assumed that hot oscillating loops have
a high temperature, 6$-$8 MK, because the Doppler shift oscillations are often observed
in the Fe~{\small XIX} and Fe~{\small XXI} lines. In this study, we directly measure the temperatures
of hot oscillating loops by analyzing coordinated SXT observations taken through two filters.
In all but one event, SXT observations are obtained around the times of oscillations
(see Figs.~\ref{fig1}-\ref{fig6}{\em d} and~\ref{fig1}-\ref{fig6}{\em e}). With the filter ratio
method \citep{tsu91, har92}, we derive the average electron temperature and emission measure for
the whole loop. Figures~\ref{fig1}-\ref{fig6}{\em f} and~\ref{fig1}-\ref{fig6}{\em g} show the
evolution of the average loop temperature and electron density. The electron density is calculated
from the emission measure by assuming the plasma depth along the line of sight to be 10 Mm (typical
SXT loop width) and a filling factor of unity. Some loops, for example Loop-5 (Fig.~\ref{fig4}), show
significant intensity variations along the loop. This may be because this is a loop
system, which is wider than 10 Mm and the density could be over-estimated.
We find that the temperature during the oscillations is in the range of 5$-$8 MK with a small variation.
Most studied loops (except for Loop-2 and Loop-4) exhibit a gradual cooling from 7$-$8 MK and the
electron density appears to vary slower than the temperature (e.g., Loop-5 in Figs.~\ref{fig4}{\em f}
and~\ref{fig4}{\em g}).
The time-averaged temperature and electron density of the loops during the oscillation are listed in
Table~\ref{tabpar}. The errors are the standard deviations from the average. Note that since SXT
missed observations of the oscillation during the main phase for event 7 (Fig.~\ref{fig6}{\em f}
and~\ref{fig6}{\em g}) and was observing during the decay phase, we take the first data point recorded
in the later decay phase as the average value. This seems reasonable because the temperature and
electron density both show a decreasing trend.

\subsection{Determination of Coronal Magnetic Field}
           \label{subcmf}

\subsubsection{Magnetic field measurement}
We have measured the oscillation period, loop length, temperature and electron density for seven
oscillating loops. With these parameters we can calculate the sound speed
($c_s=[\gamma{k_B}T/\mu{m}_p]^{1/2}$), the tube speed for the fundamental mode of a standing slow
mode wave ($c_t=2L/P$), the Alfv\'{e}n speed ($v_A=c_{s}c_t/[c_s^2-c_t^2]^{1/2}$), plasma $\beta$
($=[2/\gamma][c_s^2/v_A^2]$), and finally the mean coronal magnetic field in the loops (eq.[\ref{eqmb}]).
These parameters are listed in Table~\ref{tabpar}, and the average values and ranges of these
parameters are given in Table~\ref{tabavg}. If we estimate the period of the fundamental
mode as $\sim2L/c_s$, then its ratio to the measured period is in the range of 0.75$-$0.94 for
all seven cases, which strongly supports the scenario that the loop oscillations we have detected
are the fundamental mode.

We find Alfv\'{e}n speeds in the range 442$-$1123~km s$^{-1}$ with a mean of 830$\pm$223 km~s$^{-1}$,
consistent with the result by \citet{ofm02a} who deduced Alfv\'{e}n speeds in the range
384$-$1640 km~s$^{-1}$ with a mean of 986$\pm$385 km~s$^{-1}$ for 11 kink-mode oscillating
loops observed by TRACE. In the case of Loop-2, we have noticed that the loop system consists
of two loops, both of which intersected the slit at the region where the Doppler shift
oscillations were detected. For one of these two loops, however, we find that $c_t=2L/P=501$ km~s$^{-1}$,
larger than $c_s$. This is inconsistent with $c_t<c_s$ implied by equation~(\ref{eqpslw}). This could
be caused by the uncertainty of the loop length measurement (e.g., neither the circular nor elliptical
shape is a good approximation for this loop). It is unlikely to be an error in the loop temperature 
because the Fe~{\small XIX} line, at which loop oscillations are detected, is formed within 
a narrow temperature range.

We find the plasma $\beta$ in the range 0.15$-$0.91 with a mean of 0.33$\pm$0.26. When excluding
Loop-4 with $\beta=0.91$, $\beta$ is in the range 0.15$-$0.33 with a mean of 0.24$\pm$0.08, 
which are expected for hot ($>$6 MK) coronal loops.

Finally, we obtain the magnetic field in the oscillating loops in the range  21$-$61~G
with a mean of 34$\pm$14~G. These loops have an average height of 32$\pm$10 Mm in the range
24$-$52 Mm, which are calculated by using $h=(h_0+r)cos\theta$ (see Table~\ref{tabgeo}). 
In comparison, \citet{asc02} determined the magnetic field to be in the range 3$-$90~G from TRACE 
observations of kink-mode oscillating loops.  With two cases of kink oscillations seen by TRACE, 
\citet{nak01} inferred a narrowed range of 4$-$30~G. Recently, \citet{ver04} analyzed transverse 
oscillations of 9 coronal loops of a post-flare loop arcade 
observed by TRACE and interpreted them as standing fast kink oscillations in both the fundamental 
and second harmonics. For those loops, they obtained the magnetic field in
the range of 9$-$46~G. Our measurements are basically consistent with these results. On the other hand,
it is likely that our measurements over-estimate the magnetic field in some events due to
an over-estimate of the loop density. Measurements of TRACE loops with the temperature of
1$-$2~MK showed the electron density to be about (1$-$2)$\times10^9$ cm$^{-3}$ \citep[e.g.][]{asc99,asc00}. 
Our measurements give an average electron density of around eight times greater. The over-estimate of 
the loop density could be due to strong background emissions in the low corona, which will be
discussed in Sect.~\ref{sectbk}.

\subsubsection{Error analysis}
Now we analyze the uncertainties in the estimated magnetic field. From equation~(\ref{eqmb}) the following
formula can be derived with the error propagation law,
\begin{equation}
  \frac{\sigma_B}{B}=\sqrt{f_n^2\left(\frac{\sigma_{n}}{n}\right)^2+
                           f_T^2\left(\frac{\sigma_{T}}{T}\right)^2+
                           f_L^2\left(\frac{\sigma_{L}}{L}\right)^2+
                           f_P^2\left(\frac{\sigma_{P}}{P}\right)^2},
        \label{eqsbb}
\end{equation}
\noindent
and
\begin{mathletters}
\begin{eqnarray}
f_n &=& \left(\frac{\partial{B}}{\partial{n}}\right)\Big{/}\left(\frac{B}{n}\right)
     = \frac{1}{2},
    \label{eqfn} \\
f_T &=& \left(\frac{\partial{B}}{\partial{T}}\right)\Big{/}\left(\frac{B}{T}\right)
     = \left(\frac{C_2P^2T_6}{2L^2}-2\right)^{-1}=\frac{1}{\gamma\beta},
    \label{eqft} \\
f_L &=& \left(\frac{\partial{B}}{\partial{L}}\right)\Big{/}\left(\frac{B}{L}\right)
     = \left(1-\frac{4L^2}{C_2T_6P^2}\right)^{-1}=1+\frac{2}{\gamma\beta},
    \label{eqfl} \\
f_P &=& \left(\frac{\partial{B}}{\partial{P}}\right)\Big{/}\left(\frac{B}{P}\right)
    = f_L=1+2f_T.
    \label{eqfp}
\end{eqnarray}
\end{mathletters}
\noindent
Uncertainties in measuring the loop density, temperature, length, and oscillation periods
all contribute to the uncertainty in measurements of $B$. Equations (\ref{eqft})-(\ref{eqfp}) show
that contribution by uncertainties in density measurements is the least significant,
as $f_n$ is a constant small number. $f_T$, $f_L$ and $f_P$ all increase as
plasma $\beta$ decreases. Therefore, when $\beta$ is small, uncertainties in the magnetic field measurement
will be very sensitive to the errors in the determination of the temperature, loop length and oscillation
period. The relations in equations~(\ref{eqft})-(\ref{eqfp}) also imply that for relatively high $\beta$
loops, i.e., coronal loops with high temperature, high density, and weak magnetic field,
there is less relative uncertainty in the estimated magnetic field, because of a stronger coupling between
the perturbations of magnetic pressure and thermal pressure.

The errors of oscillation period measurement listed in Table~\ref{tabosc} are given by the least-square fit.
The errors of temperature and electron density measurements listed in Table~\ref{tabpar} are the
standard deviations of the time average during the oscillation. Estimates of the uncertainty for the
loop length measurement is difficult, because it involves uncertainties in resolving fine loop structures,
the loop shape, and locating loop footpoints. For convenience, we assume the relative
error of loop length measurement ($\sigma_{L}/L$) as 5\%, which is comparable to those of temperature and 
electron density measurements ($\sigma_T/T$ and $\sigma_n/n$). By applying equation~(\ref{eqsbb})
the errors of the magnetic field $B$ are calculated and listed in Table~\ref{tabpar}. For the 7 analyzed 
loops, we obtain the relative error of the magnetic field $B$ to be 40\% on average with $\bar{f_T}\approx2.4$ 
and $\bar{f_L}=\bar{f_P}\approx5.9$. Given the same set of $f_T$, $f_L$ and $f_P$ values, the
relative errors of $n_e$, $T$, $L$ and $P$ have to be within 10\%, 2\%, 1\% and 1\%,
respectively, in order to improve the accuracy of $B$ measurement to 10\%. This suggests that
the improvement of measurement accuracies in $T$, $L$ and $P$ is essential. In addition, we estimate
the uncertainty in $L$ based on the inequality, $c_t<c_s$. Since all 7 cases show  
$2L_{obs}/P=2(L+\sigma_{L})/P<c_s$ if only considering the uncertainty in $L$, we can obtain 
$\sigma_{L}/L<(1+\gamma\beta/2)^{1/2}-1$. Given the upper limit of $\beta=0.33\pm0.26$, 
yielding $\sigma_{L}/L<22\%$.  This implies that the uncertainty in $L$ in our measurements 
is not larger than 22\%, e.g. caused by a departure of loops from the circular shape we have assumped. 

We note that the mean density and temperature are used to determine $B$, while $T$ and $n$
both evolve with time because of loop heating and cooling. For more accurate determination of
$B$, not only measurements of $T$ and $n$ should be conducted with improved accuracy, but also
their temporal variations should be delineated from the uncertainty estimates. Theoretically, the evolving
$T$ and $n$ may lead to a time dependent loop oscillation period, given that the coronal magnetic field
and loop length are nearly constant (or their evolution time scales are much longer than
the cooling time) during the oscillation. Hopefully, high cadence observations of time series
of Doppler shift oscillations by future missions will allow us to examine the time profiles
of the oscillation period, $P(t)$, with respect to $T(t)$ and $n(t)$. If variations of these three parameters
follow equation~(\ref{eqmb}), we can minimize uncertainties in determining $B$ caused by using the time averaged
temperature and electron density.

\subsubsection{Effects of the Background Subtraction}
\label{sectbk}
In addition, we consider the effects of the background plasma emission. Plasma emissions outside oscillating
loops along the line of sight could cause an over-estimate of the electron
density in the loop (assuming a filling factor of unity). However, to make a background subtraction
for the whole loop is not an easy matter. An active region loop is rarely an isolated loop, and
often overlaps with nearby loops along the line-of-sight. This may cause a local brightening along the loop
(e.g., Loop-5 shown in Fig.~\ref{fig4}). In this case we have no way to correct the emissions from other
loops. Here we only consider effects of mean background emission in regions adjacent to the oscillating loops
in the FOV of the observations (e.g., a small box between Loop-1 and Loop-2 in Fig.~\ref{fig1}{\em a}). With this
mean background emission subtracted, we find a slight ($\sim3$\%) increase in temperature in most cases and
a decrease in electron density for all loops by about 12\% (e.g., Figs.~\ref{fig1}{\em f} and~\ref{fig1}{\em g}).
We list the time-averaged temperatures and electron densities, the derived sound speeds, Alfv\'{e}n speeds, plasma
$\beta$ and magnetic field in Table~\ref{tabpbk} and their averages and ranges in Table~\ref{tabavg}.
On average, the background subtraction causes a decrease in the Alfv\'{e}n speed by about 6\%, an increase of
the plasma $\beta$ by about 15\% and a decrease in the magnetic field by about 9\%.  The effect is not significant
compared with uncertainties discussed above, specifically uncertainties in the loop length,
as background emission mainly leads to uncertainties in electron density, which has a relatively small
influence on $B$ compared to the other parameters.

We note that the above-mentioned background emission is obtained from a region above the oscillating loop. 
Emission in regions below the loops is much stronger (e.g., the cases of Loops 5, 6 and 7).
For a comparison, we take the background emission from a small box below the loop as the upper limit of
the background emission. We repeat the above calculations and find that on average the temperature rises 
by about 17\% and the electron density reduces by about 40\%. Both lead to a reduction of the magnetic 
field. The combined effect leads to an estimate of the magnetic fields with a mean of 22$\pm$13~G in 
the range 12$-$51~G. The magnetic fields are reduced by about 35\% compared to the values obtained before 
the background subtraction. Therefore, an accurate measurement of coronal loop temperature and density is 
crucial to the determination of magnetic field.

\section{Conclusions and Discussions}

This study presents the first effort to determine mean magnetic field in coronal loops exhibiting
standing slow mode oscillations. We analyzed seven Doppler shift oscillation events observed by
SUMER in the Fe~{\small XIX} line that have coordinated SXR imaging observations  obtained
by SXT in at least two filters. This has allowed us to measure the oscillation periods of
identified loops, and determine the loop geometrical and physical parameters such as the loop length, 
plasma temperature and electron density. Magnetic field is calculated from the loop plasma and 
oscillation parameters. We summarize our results as follows:

1. It is evident that all oscillation events in this study exhibit the fundamental mode of standing slow
waves, as pronounced Doppler shift oscillations were located at the loop apex, and the oscillation
period is approximately twice the sound wave transit time along the loop.

2. With the filter ratio method, we obtain from the SXT images the temperature and electron density
of the oscillating loops. The temperatures are in the range of 5$-$8 MK, consistent with the fact
that the Doppler shift oscillations are preferentially observed in the hot Fe~{\small XIX} line.
The characteristics of temperature and electron density evolution agree with results from numerical
simulations with an impulsive heating function \citep{war03,pat05}.

3. Applying MHD wave theory of standing slow mode waves, we obtain for the seven oscillating loops
Alfv\'{e}n speeds in the range 442$-$1123 km~s$^{-1}$ with a mean of 830$\pm$223 km~s$^{-1}$
and plasma $\beta$s in the range 0.15$-$0.91 with a mean of 0.33$\pm$0.26. The mean magnetic field
in the oscillating loops is estimated to be in the range 21$-$61~G with a mean of 34$\pm$14~G. Determination
of $B$ is sensitive to uncertainties in measurements of temperature, loop length and oscillation period.
The magnetic field can be derived with higher accuracy in loops with a higher $\beta$. Background
subtraction in these events could lead to a decrease in the derived magnetic field by 9\%$-$35\%.

Measurements of plasma properties in coronal loops will be greatly improved with the upcoming space missions.
Solar TErrestrial RElations Observatory (STEREO) will directly measure the loop geometry in 3-D, 
yielding a more accurate loop length measurement.
The EUV Imaging Spectrometer (EIS) on Solar-B is capable of quick raster scanning to observe the
Doppler shift oscillation of active region coronal loops both off the limb and on the disk.
The two EUV bands, 170$-$210 \AA~and 250$-$290 \AA~contain many coronal and flare lines formed in
a wide temperature range, 1$-$20 MK, which can provide both temperature and density diagnostics using
the line pair ratio method and a differential emission measure (DEM) analysis with several lines. Some
studies have indicated that active region loops may have multi-thermal components along the
line of sight \citep[e.g.][]{har92,sch01,sch06}, so temperature measurements through filter and
line ratios are possibly inadequate \citep{mar02}. Instead, we should use the
DEM-weighted temperature in our analysis. An impulsively generated flare disturbance could
excite both global fast kink-mode and slow-mode oscillations in coronal loops as shown by MHD
simulations \citep{sel06}. With joint observations by the X-Ray Telescope (XRT) and
EIS on Solar-B such events in hot coronal loops could be captured, allowing a direct comparison of the estimated
magnetic field from both kinds of oscillations. Although the slow mode wave has weak connection with
the coronal magnetic field compared to the fast kink mode wave, an evident advantage
is that the occurrence rate of the slow mode oscillations is much higher than that of the fast kink
mode oscillations  \citep[e.g.][]{wan03b}, which may make the former
a more useful diagnostic tool.

From the theoretical point of view, it is desirable to develop a non-uniform, cylindrical magnetic loop model
including curvature and stratification effects (e.g. with longitudinal temperature, density
and magnetic structures) in order to describe the MHD wave properties in a more realistic model, thus
can guide or be adapted to the new discoveries likely to emerge from new observations, e.g., the varied
wave properties along the loop and high harmonics. These new observations and theories will lead to significant
progress in {\em coronal seismology} and help us obtain the distribution of magnetic field and
other physical parameters along the loop, which may serve to solve problems such as coronal
heating and solar wind acceleration.

\acknowledgments
We would like to thank the referee, Valery Nakariakov, for his valuable comments and suggestions. 
SUMER is financially supported by DLR, CNES, NASA and the ESA PRODEX programme
(Swiss contribution). SOHO is a project of international co-operation between
ESA and NASA. $Yohkoh$ is a mission of the Institute of Space and Astronautical
Science (Japan), with participation from the US and UK. This work was supported 
by NASA grant NNG06GA37G.

 \clearpage

 \begin{figure}
 \plotone{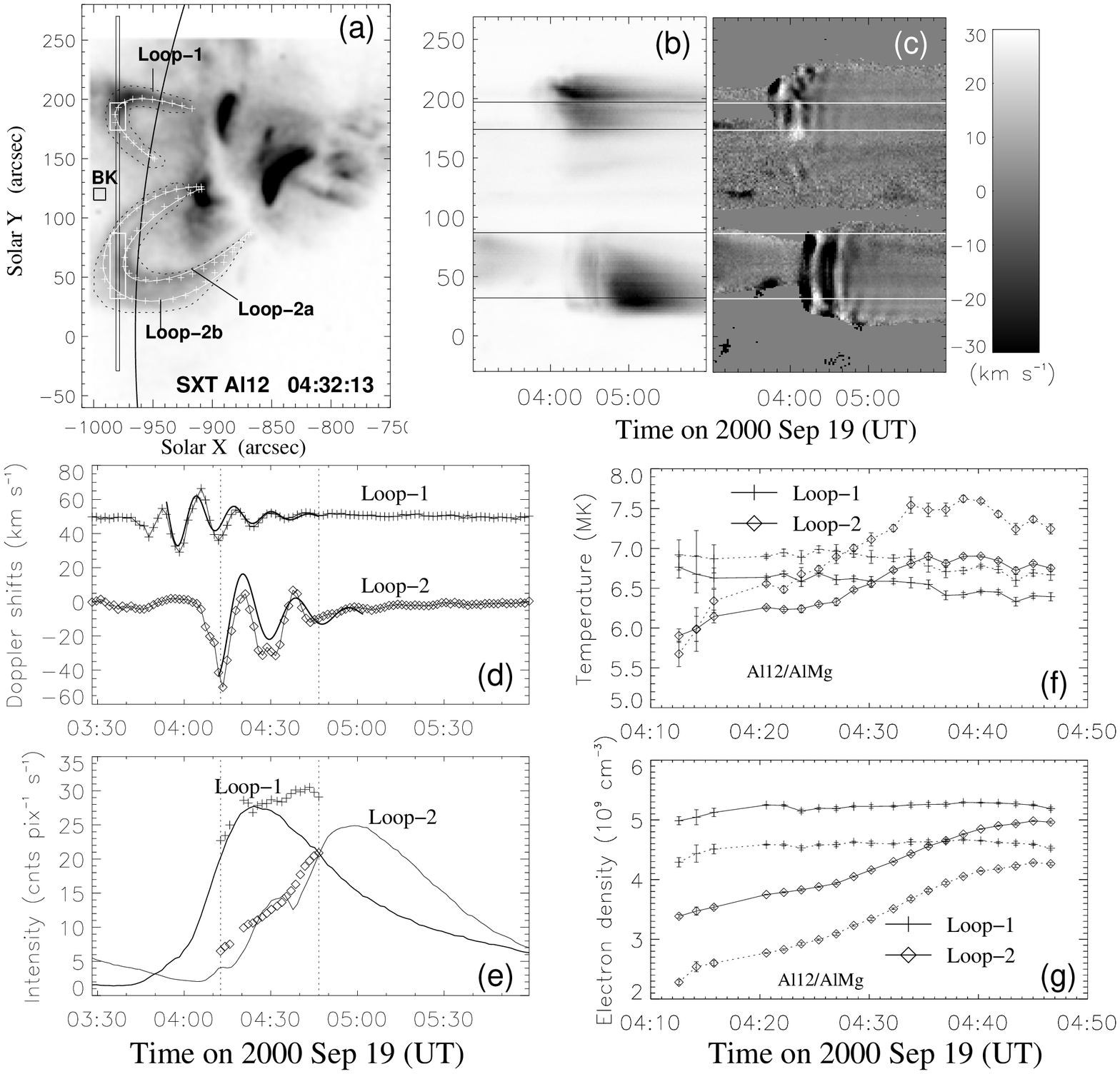}
 \caption{\label{fig1}
  Oscillation events No.1 and No.2 on 19 September 2000. (a) The oscillating soft X-ray loops
 (outlined with {\em diamonds}) fitted with a circular model (for Loop-1 and Loop-2a) and
an elliptical model (for Loop-2b). The SUMER spectrometer slit position is indicated as a vertical
black box. (b) Line-integrated intensity and (c) Doppler shift time series in the Fe~{\small XIX}
line at a fixed slit position. (d) Average time profiles of Doppler shifts along the cuts shown
in (c). The thick solid curves are the best fit to a damped sine function. (e) Average time
profiles of line-integrated intensity along the cuts shown in (b). The overlaid data points
({\em crosses} and {\em diamonds}) are the soft X-ray intensity (in arbitrary units) of SXT
AlMgMn filter averaged in the white boxes shown in (a). (f) Temperature and (g) electron density
light curves (solid lines) calculated for the whole loop (outlined with the dotted curves in (a))
with the filter ratio method. The dotted curves represent the case after the subtraction of the
background emission, which is taken as the average in a black box, marked $BK$.}
 \end{figure}

 \begin{figure}
 \plotone{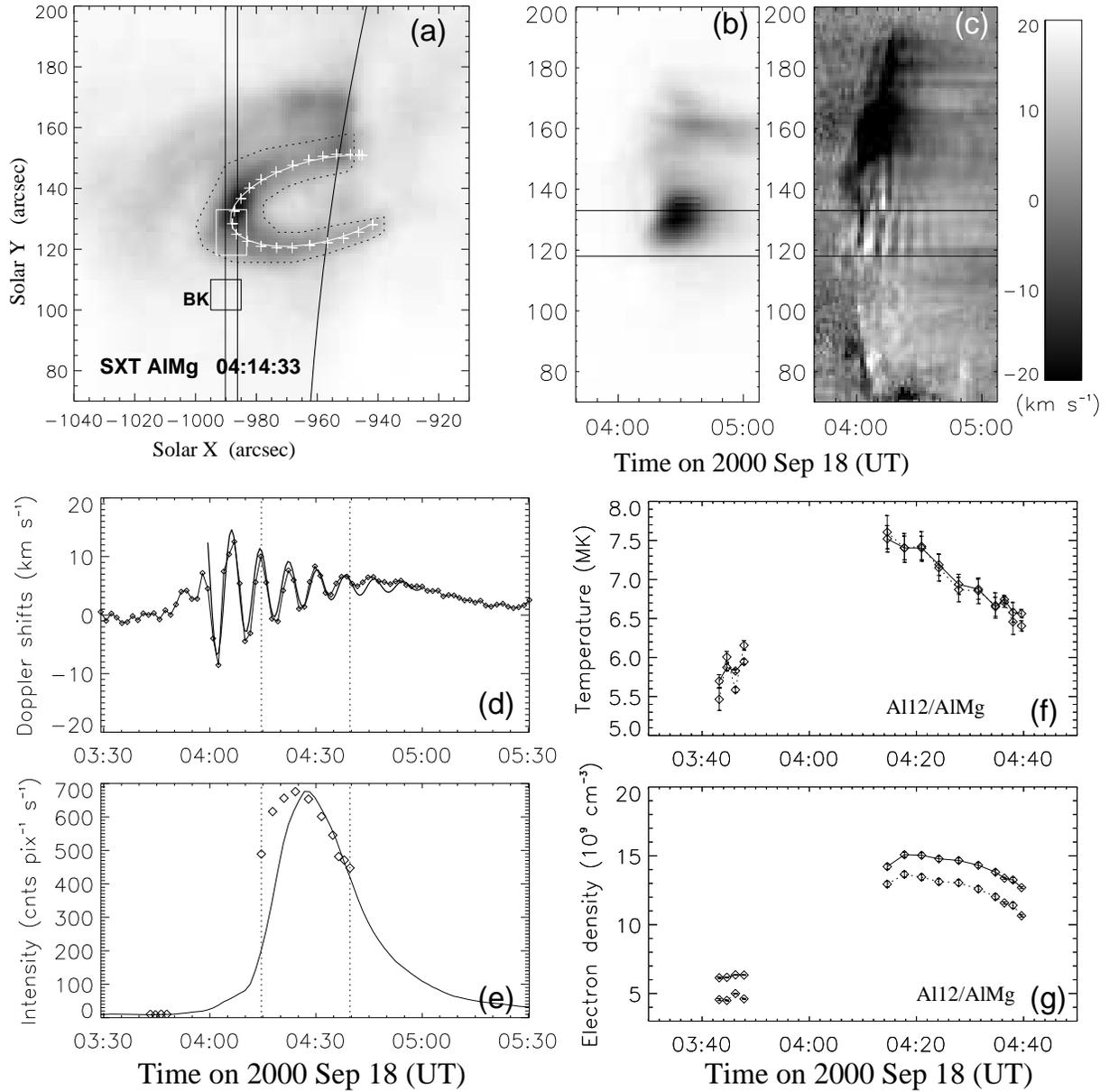}
 \caption{\label{fig2}
  Oscillation event No.3 on 18 September 2000.}
\end{figure}

 \begin{figure}
 \plotone{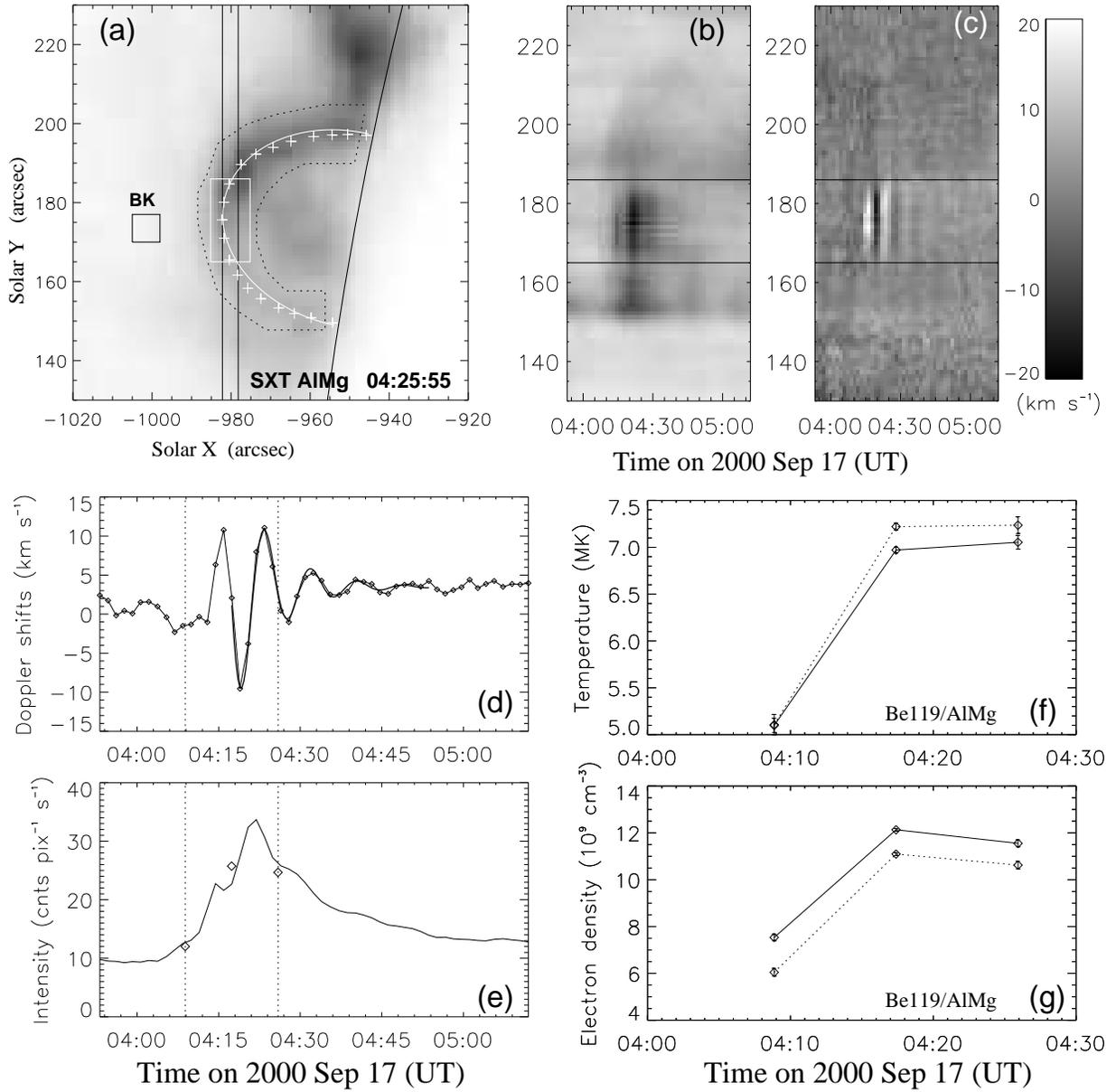}
 \caption{\label{fig3}
 Oscillation event No.4 on 17 September 2000, 04:15.}
 \end{figure}

  \begin{figure}
    \plotone{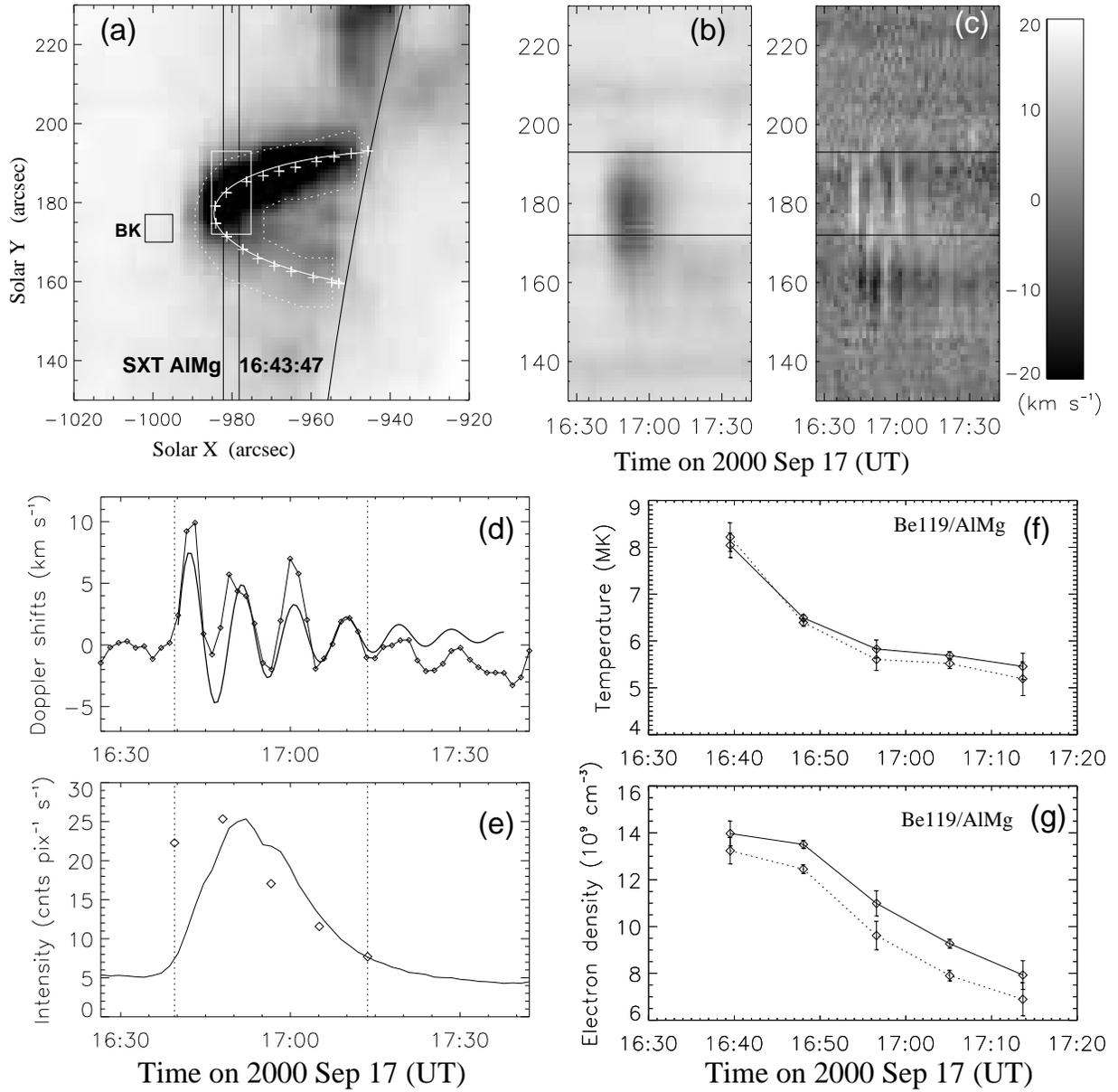}
  \caption{\label{fig4}
 Oscillation event No.5 on 17 September 2000, 16:40.}
  \end{figure}

 \begin{figure}
   \plotone{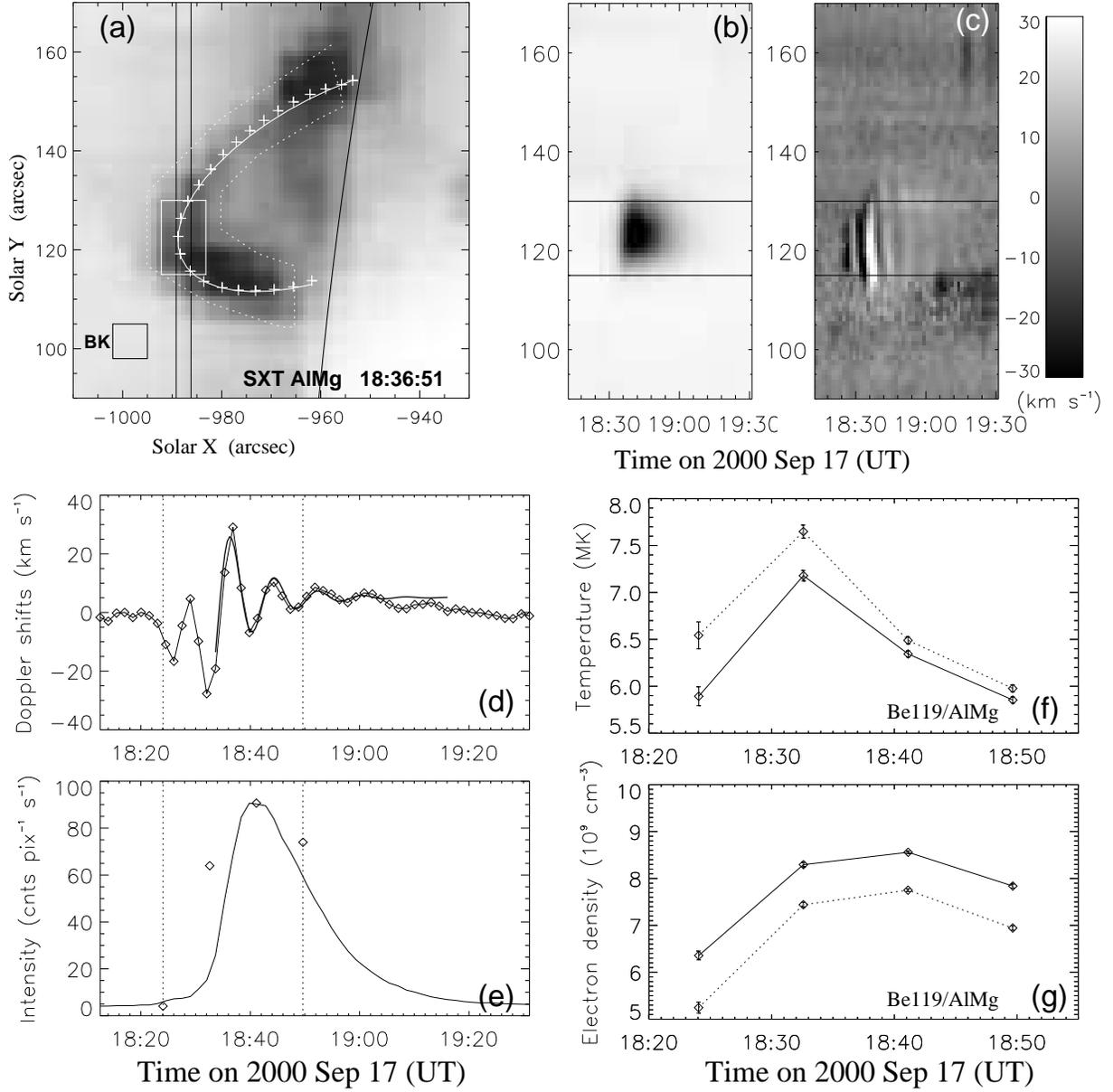}
      \caption{\label{fig5}
 Oscillation event No.6 on 17 September 2000, 18:30.}
 \end{figure}

  \begin{figure}
    \plotone{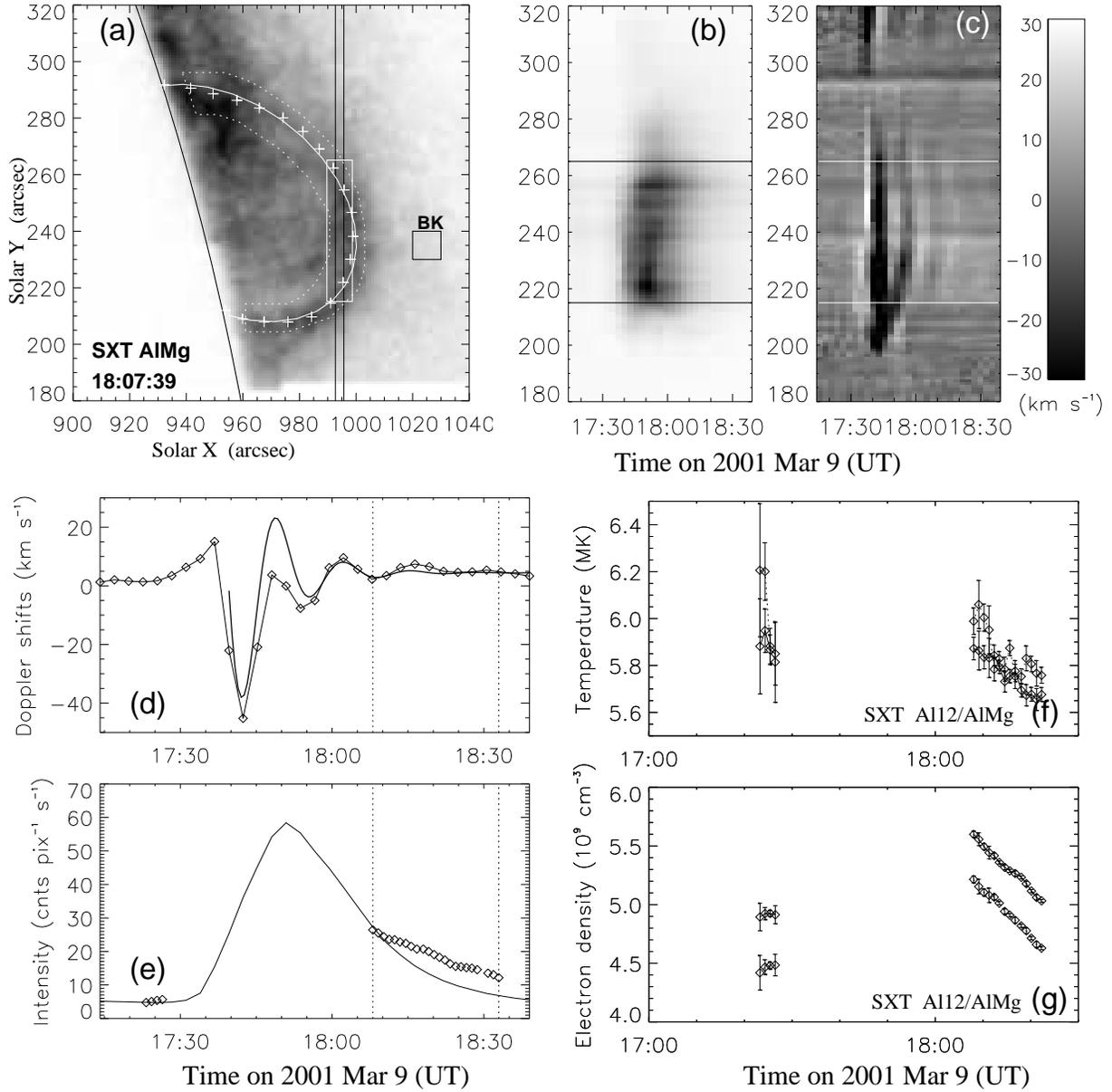}
       \caption{\label{fig6}
 Oscillation event No.7 on 9 March 2001.}
\end{figure}

  \begin{figure}
    \plotone{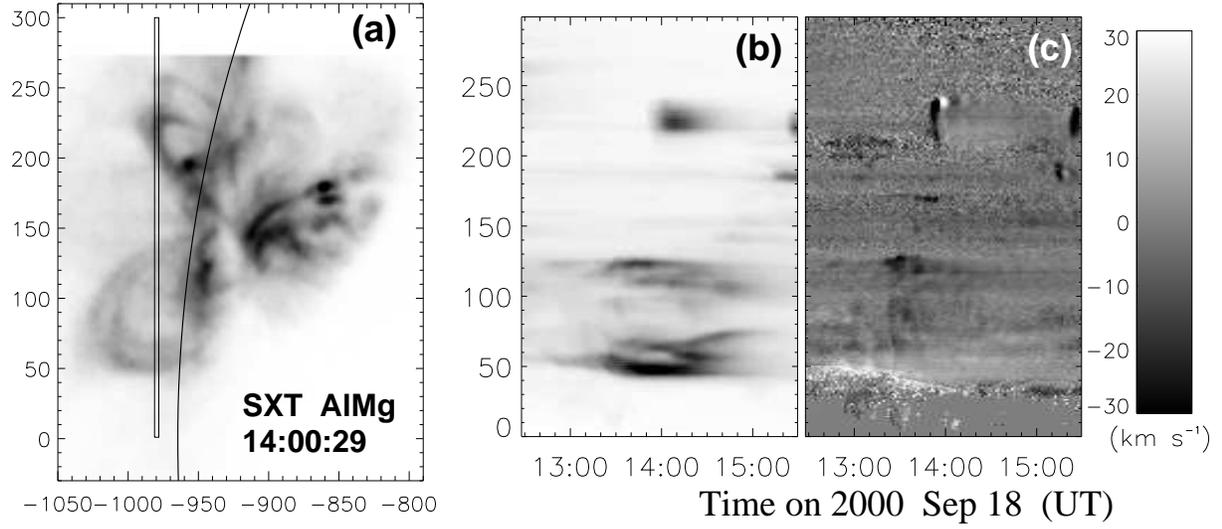}
       \caption{\label{fig7}
  Loop brightening event on 18 September 2000. (a)-(c) Same as in Fig.~\ref{fig1}. }
\end{figure}
 \clearpage

 \begin{deluxetable}{llrrrrrl}
 \tablecaption{Time series analysis of Doppler shift oscillation\tablenotemark{a}.   \label{tabosc}}
 \tablewidth{0pt}
 \tablehead{
 \colhead{Loop} &  \colhead{$t_{\rm 0}$} & \colhead{$V_{\rm m}$} & \colhead{$P$} &
 \colhead{$\phi$} & \colhead{$\tau_{\rm d}$} & \colhead{A} & \colhead{$N_{\rm P}$} \\
 \colhead{No.}  &      & \colhead{(km s$^{-1}$)} & \colhead{(min)} & \colhead{(rad)} &
 \colhead{(min)} & \colhead{(Mm)} & }
 \startdata
  1 & 03:53:53 19-Sep-00 & 21\hspace{5mm} & 12.9$\pm$0.2\hspace{5mm} & 2.73 & 18.1$\pm$3.1 & 2.64 & 4 \\
  2 & 04:11:54 19-Sep-00 & 36\hspace{5mm} & 18.3$\pm$0.7\hspace{5mm} & $-$1.47 & 19.9$\pm$4.2 & 6.28 & 2 \\
  3 & 03:59:25 18-Sep-00 & 14\hspace{5mm} & 8.1$\pm$0.1\hspace{5mm}  & 2.56    & 19.8$\pm$5.0 & 1.05 & 5 \\
  4 & 04:17:25 17-Sep-00 & 17\hspace{5mm} & 8.5$\pm$0.2\hspace{5mm} & $-$3.00  & 7.4$\pm$1.7  & 1.33 & 4 \\
  5 & 16:40:09 17-Sep-00 & 8\hspace{5mm}  & 9.3$\pm$0.3\hspace{5mm} & 0.12     & 19.1$\pm$9.2 & 0.68 & 5 \\
  6 & 18:33:39 17-Sep-00 & 30\hspace{5mm} & 8.1$\pm$0.3\hspace{5mm} & $-$0.66  & 7.2$\pm$1.8 & 2.30 & 3 \\
  7 & 17:39:37 09-Mar-01 & 61\hspace{5mm} & 13.3$\pm$0.7\hspace{5mm} & $-$3.04 & 8.1$\pm$2.3 & 7.44 & 2\\
 \enddata
\tablenotetext{a}{ $t_0$ is the start time of the modelled time series, $V_{\rm m}$ the Doppler
velocity amplitude derived by the best fit of a damped sine-function to the oscillations,
$P$ the oscillation period, $\phi$ the phase of the oscillation,
$\tau_{\rm d}$ the decay time ($\tau_{\rm d}=1/\lambda$), $A$ the displacement amplitude
(defined as $A=V_{\rm m}/(\omega^2+\lambda^2)^{1/2}$), $N_{\rm P}$ the
number of periods over which an oscillation was detected.}
\end{deluxetable}

\begin{deluxetable}{lrrrrrrrr}
 \tablecaption{Geometrical parameters of oscillating loops derived from Yohkoh/SXT
   images\tablenotemark{a}.   \label{tabgeo}}
 \tablewidth{0pt}
 \tablehead{
  \colhead{Loop} &  \colhead{$l_{\rm 0}-l_{\rm \sun}$} & \colhead{$b_{\rm 0}-b_{\rm \sun}$} &
  \colhead{$\alpha$} & \colhead{$\theta$} & \colhead{$\psi$} & \colhead{h$_{\rm 0}$} &
  \colhead{$r$} & \colhead{$L$} \\
  \colhead{No.}  &  \colhead{(deg)} & \colhead{(deg)} & \colhead{(deg)} & \colhead{(deg)} &
  \colhead{(deg)} & \colhead{(Mm)} & \colhead{(Mm)} & \colhead{(Mm)} }
 \startdata
 1 & $-$80.2\hspace{5mm} & 10.2\hspace{5mm} & 16.8 & 17.6 & 20.9 & $-$22.3 & 57.8 & 134 \\
 2a & $-$68.2\hspace{5mm} & 6.3\hspace{5mm} & $-$17.0 & $-$42.1 & 25.9 & 20.5 & 50.0 & 199 \\
 2b\tablenotemark{b} & $-$68.3\hspace{5mm} & 6.3\hspace{5mm} & $-$18.5 & $-$44.0 & 26.7 & 35.2 & 70.3/46.4 & 275 \\
 3 & $-$81.4\hspace{5mm} & 8.3\hspace{5mm} & $-$26.7 & $-$24.0 & 27.0 & 11.4 & 21.6 & 92 \\
 4 & $-$90\hspace{5mm} & 10.3\hspace{5mm} & 48.8 & 2.9 & 48.8 & 0.24 & 23.5 & 74 \\
 5 & $-$90\hspace{5mm} & 10.5\hspace{5mm} & 18.0 & 7.0 & 18.0 & $-$16.9 & 42.3 & 98 \\
 6 & $-$90\hspace{5mm} & 8.0\hspace{5mm} & 33.0 & $-$30.2 & 33.0 & $-$1.9 & 27.2 & 82 \\
 7 & 90\hspace{5mm} & 16.5\hspace{5mm} & 57.9 & 35.5 & 57.9 & 9.0 & 37.0 & 135 \\
 \enddata
\tablenotetext{a}{$l_0-l_{\sun}$ and $b_0-b_{\sun}$ are the heliographic
longitude and latitude relative to Sun center for the midpoint of the loop
footpoint baseline. $\alpha$ is the azimuth angle of the loop baseline
to the east-west direction. $\theta$ is the inclination angle of the loop plane to
the vertical. $\psi$ is the angle between the loop baseline and line-of-sight.
$h_0$ is the height of the circular loop center in the loop plane. $r$ is the radius
of the circular loop. $L$ is the loop length.}
\tablenotetext{b}{To get the best fit, an elliptical shape is assumed for this loop.
The two values in $r$ are the semi-major and semi-minor axis lengths, where the semi-major
axis is parallel to the solar surface.}
\end{deluxetable}

\begin{deluxetable}{crrccrrr}
 \tablecaption{Temperature, electron density, sound speed, tube speed, Alfv\'{e}n
speed, plasma $\beta$ and the mean magnetic field inside oscillating loops.  \label{tabpar}}
 \tablewidth{0pt}
 \tablehead{
 \colhead{Loop} & \colhead{$T$} & \colhead{$n_{\rm e}$} & \colhead{$c_{\rm s}$} &
 \colhead{$c_{\rm t}$} & \colhead{$v_{\rm A}$} & \colhead{$\beta$} & \colhead{$B$\tablenotemark{c}} \\
 \colhead{No.}  & \colhead{(MK)} & \colhead{($10^9$cm$^{-3}$)} & \colhead{(km s$^{-1}$)} &
 \colhead{(km s$^{-1}$)} & \colhead{(km s$^{-1}$)} & & \colhead{(G)} }
 \startdata
1 & 6.6$\pm$0.1 & 5.2$\pm$0.1\hspace{5mm}  & 390 & 346 & 749\hspace{5mm}  & 0.33 & 25$\pm$6 \\
2\tablenotemark{a} & 6.5$\pm$0.3 & 4.3$\pm$0.5\hspace{5mm} & 388 & 362 & 1025\hspace{5mm} & 0.17 & 31$\pm$16\\
3 & 7.0$\pm$0.3 & 14.1$\pm$0.8\hspace{5mm} & 402 & 379 & 1123\hspace{5mm} & 0.15 & 61$\pm$30 \\
4 & 6.4$\pm$0.9 & 10.4$\pm$2.0\hspace{5mm} & 385 & 290 & 442\hspace{5mm}  & 0.91 & 21$\pm$4 \\
5 & 6.3$\pm$0.9 & 11.1$\pm$2.3\hspace{5mm} & 382 & 351 & 900\hspace{5mm}  & 0.22 & 43$\pm$25 \\
6 & 6.3$\pm$0.5 & 7.8$\pm$0.9\hspace{5mm}  & 382 & 337 & 723\hspace{5mm}  & 0.33 & 29$\pm$10 \\
7 & 5.9$\pm$0.3\tablenotemark{b} & 5.6$\pm$0.5\tablenotemark{b}\hspace{3mm} & 369 & 338 & 845\hspace{5mm} & 0.23 & 29$\pm$14 \\
\enddata
\tablenotetext{a}{In calculations of $c_t$, $v_A$, $\beta$ and $B$, the loop length was taken from
Loop-2a in Table~\ref{tabgeo}.}
\tablenotetext{b}{In this case $T$ and $n_e$ are taken as the maximum values because SXT missed observations
of the event during the main phase.}
\tablenotetext{c}{In calculations of the uncertainty in $B$, the uncertainty in the loop length
is taken as $5\%L$.}
\end{deluxetable}

\begin{deluxetable}{crrccrrr}
 \tablecaption{Same as Table~\ref{tabpar} but for the cases with the background subtraction.  \label{tabpbk}}
 \tablewidth{0pt}
 \tablehead{
 \colhead{Loop} & \colhead{$T$} & \colhead{$n_{\rm e}$} & \colhead{$c_{\rm s}$} &
 \colhead{$c_{\rm t}$} & \colhead{$v_{\rm A}$} & \colhead{$\beta$} & \colhead{$B$\tablenotemark{a}} \\
 \colhead{No.}  & \colhead{(MK)} & \colhead{($10^9$cm$^{-3}$)} & \colhead{(km s$^{-1}$)} &
 \colhead{(km s$^{-1}$)} & \colhead{(km s$^{-1}$)} & & \colhead{(G)} }
 \startdata
1 & 6.8$\pm$0.1 & 4.6$\pm$0.1\hspace{5mm}  & 396 & 346 & 711\hspace{5mm}  & 0.37 & 22$\pm$5 \\
2 & 7.0$\pm$0.5 & 3.4$\pm$0.6\hspace{5mm}  & 402 & 362 & 837\hspace{5mm}  & 0.28 & 22$\pm$9 \\
3 & 7.0$\pm$0.4 & 12.4$\pm$0.9\hspace{5mm} & 402 & 379 & 1123\hspace{5mm} & 0.15 & 57$\pm$29 \\
4 & 6.5$\pm$1.0 & 9.3$\pm$2.3\hspace{5mm}  & 388 & 290 & 438\hspace{5mm}  & 0.94 & 19$\pm$4 \\
5 & 6.2$\pm$1.1 & 10.0$\pm$2.5\hspace{5mm} & 378 & 351 & 943\hspace{5mm}  & 0.19 & 43$\pm$31 \\
6 & 6.7$\pm$0.6 & 6.8$\pm$1.0\hspace{5mm}  & 393 & 337 & 656\hspace{5mm}  & 0.43 & 25$\pm$7 \\
7 & 6.1$\pm$0.3 & 5.2$\pm$0.6\hspace{3mm}  & 375 & 338 & 781\hspace{5mm}  & 0.28 & 26$\pm$10 \\
\enddata
\tablenotetext{a}{In calculations of the uncertainty in $B$, the uncertainty in the loop
length is taken as $5\%L$.}
\end{deluxetable}

\begin{deluxetable}{lrrcrr}
 \tablecaption{Average and range of physical parameters of 7 oscillating loops for the
cases before and after subtractions of the background emission ($I_{bk}$). \label{tabavg}}
 \tablewidth{0pt}
 \tablehead{\colhead{Parameter}  & \multicolumn{2}{c}{Before $I_{bk}$ subtraction}& & \multicolumn{2}{c}{After $I_{bk}$ subtraction} \\
  \cline{2-3}  \cline{5-6}\\
   & \colhead{Average} & \colhead{Range}& & \colhead{Average} & \colhead{Range}}
 \startdata
  Oscillation period $P$ (min)        & 11.2$\pm$3.8  & 8.1$-$18.3 & &\multicolumn{2}{c}{same} \\
  Loop length $L$ (Mm)    & 116$\pm$44     & 74$-$199  & & \multicolumn{2}{c}{same} \\
  Temperature $T$ (MK)  & 6.4$\pm$0.3    & 5.9$\pm$7.0& & 6.6$\pm$0.4 & 6.1$-$7.0 \\
  Electron density $n_e$ ($10^9$ cm$^{-3}$) & 8.4$\pm$3.6 & 4.3$\pm$14.1& & 7.4$\pm$3.3 & 3.4$-$12.4 \\
  Sound speed $c_s$ (km s$^{-1}$) & 385$\pm$10   & 369$-$402& & 391$\pm$11 & 375$-$402 \\
  Alfv\'{e}n speed $v_A$ (km s$^{-1}$) & 830$\pm$223 & 442$-$1123& & 784$\pm$218 & 438$-$1123 \\
  Plasma $\beta$ & 0.33$\pm$0.26  & 0.15$-$0.91& & 0.38$\pm$0.27 & 0.15$-$0.94 \\
  Magnetic field $B$ (G) & 34$\pm$14 & 21$-$61& & 31$\pm$14 & 19$-$57\\
 \enddata
\end{deluxetable}
\end{document}